\pdfoutput=1
\documentclass[12pt]{article}
\usepackage{fullpage}
\usepackage{pdfpages}
\usepackage[pdftex,bookmarks=true]{hyperref}
\usepackage{longtable}

\hypersetup{
  bookmarksnumbered=true,
  bookmarksopen=true,
  pdftitle={Lecture Notes on Network Information Theory},
  pdfauthor={Abbas El Gamal and Young-Han Kim}
}

\begin{document}
\title{\sffamily Lecture Notes on Network Information Theory}
\author{\sffamily Abbas El Gamal and Young-Han Kim}
\date{}
\maketitle

\noindent
These lecture notes have been converted to a book titled {\em Network
Information Theory} published recently by Cambridge University
Press. This book provides a significantly expanded exposition of the
material in the lecture notes as well as problems and bibliographic
notes at the end of each chapter.
The authors are currently preparing a set of slides based on the book
that will be posted in the second half of 2012.

More information about the book can be found at
\begin{quote}
http://www.cambridge.org/9781107008731/.
\end{quote}
The previous (and obsolete) version of the lecture notes can be found at
\begin{quote}
http://arxiv.org/abs/1001.3404v4/.
\end{quote}
\end{document}